\documentclass[12pt]{article}
\usepackage{graphicx}
\usepackage{epsf}

\def\f{\frac}

\def\n{\nu}
\def\m{\mu}

\def\t{\theta}
\def\p{\partial}

\def\be{\begin{equation}}
\def\ee{\end{equation}}
\def\bea{\begin{eqnarray}}
\def\eea{\end{eqnarray}}
\def\a{\alpha}
\def\ba{\begin{array}}
\def\ea{\end{array}}

\def\l{\left}
\def\r{\right}

\begin{document}
\begin{titlepage}
\begin{center}
{\Large \bf William I. Fine Theoretical Physics Institute \\
University of Minnesota \\}
\end{center}
\vspace{0.15in}
\begin{flushright}
FTPI-MINN-10/04 \\
UMN-TH-2834/10 \\
January 2010 \\
\end{flushright}
\vspace{0.15in}
\begin{center}
{\Large \bf Semiclassical Calculation of Photon-Stimulated Schwinger Pair Creation   \\}
\vspace{0.15in}
{\bf A. Monin \\}
School of Physics and Astronomy, University of Minnesota, \\ Minneapolis, MN
55455, USA, \\
and \\
{\bf M.B. Voloshin  \\ }
William I. Fine Theoretical Physics Institute, University of
Minnesota,\\ Minneapolis, MN 55455, USA \\
and \\
Institute of Theoretical and Experimental Physics, Moscow, 117218, Russia
\\[0.1in]
\end{center}

\begin{abstract}
We consider the electron-positron pair creation by a photon in an external constant electric field.
The presented treatment is based on a purely quasiclassical calculation of the imaginary part of the on-shell photon polarization operator. By using this approach we find the pair production rate for photons with polarization parallel as well as orthogonal to the external electric field in the leading order in the parameter $eE / m ^ 2$, which has been recently found by other methods. For the orthogonal polarization we also find a new contribution to the rate, which is leading in the ratio of the photon energy to the electron mass $\omega/m$. We also reproduce by a purely geometrical calculation the exponential factor in the probability of the stimulated pair creation at arbitrary energy of the photon. 
\end{abstract}

\end{titlepage}
\newpage

\section{Introduction.}
One of the well known problems in QED is that  of spontaneous creation of electron-positron pairs in an external electromagnetic field. This phenomenon, predicted theoretically long ago \cite{Sauter,Heisenberg:1935qt} and conventionally called a Schwinger process~\cite{Schwinger}, is greatly suppressed for practically achievable strength $E$ of the electric field.   Indeed the rate of pair creation in a field that is less than the critical one $E_{\rm crit}= m^2/e \sim 10^{16}$\, V/cm is exponentially suppressed. Namely the rate of pair creation per unit time and in unit volume $V$ is given by 
\be
{\Gamma \over V} = \f {( e E ) ^ 2} {4 \pi ^3} \sum _ {n = 1} ^ { + \infty } \f {1} {n ^ 2} \exp \l ( - \f {\pi \, m ^ 2 } {e E} \, n \r )~.
\label{ehs}
\ee
It has been recently suggested~\cite{dgs2008}, that the creation rate can be enhanced by superposing a weak beam of energetic photons with a strong, but low-frequency electric field, that can be created by lasers. The slow laser pulse can be treated as a constant electric field, and the problem is thus reduced to that of a photon-stimulated Schwinger pair creation.  The appropriate quantity characterizing the effect of the photon on the pair production can be formulated as the attenuation rate $\kappa$ for the photon propagation, so that the intensity of the photon beam decreases at length $L$ as $\exp (- \kappa L)$ due to the photon absorption by creation of the $e^+e^-$ pairs. The attenuation rate, in turn, can be related to the on-shell imaginary part of the vacuum polarization operator $\Pi$ in external electric field~\cite{dgs2009}:
\be
\kappa = -{1 \over \omega} \, \Im \Pi~.
\label{kapi}
\ee 
A calculation of the attenuation rate for the photons based on this formula has been recently performed~\cite{dgs2009} using an integral representation~\cite{Dittrich:2000zu} for the exact one-loop QED vacuum polarization in an external field. In this way explicit formulas were found at small energy of the photon $\omega/m \ll 1$, as well as the behavior of the exponential factor in the pair production rate at arbitrary $\omega$. A similar result in the small $\omega$ limit has been also found~\cite{mv2009} directly in terms of $\kappa$ by using a thermal bath approach~\cite{mvsp,mvsw} and applying a semiclassical treatment to the Schwinger process at a small but finite temperature.

In the present paper we consider the on-shell imaginary part of the vacuum polarization operator in external field within a purely semiclassical approach not using the integral form~\cite{Dittrich:2000zu}. We believe that such an approach provides some additional insight in the physical origin of various terms describing the photon-stimulated pair production, which may be useful in further studies of this subject. In terms of the specific results in this paper we reproduce the formulas of Refs.~\cite{dgs2009, mv2009}, and additionally calculate a contribution to the attenuation rate arising from the electron magnetic moment, which contribution dominates, in a situation where $\omega^2$ is larger than $e E$, over that recently found~\cite{dgs2009} for photons polarized orthogonally to the external field.

There are two dimensionless parameters in the considered problem (apart from fine structure constant $\a = e^ 2 /4 \pi$), namely
$eE / m ^ 2$ and $\omega / m$. The polarization operator can be split into two parts $\Pi(q) = \Pi _ \| + \Pi _ \bot$, corresponding to the photons having their polarization along or orthogonal to the field. In the leading order in $eE / m ^ 2$ only the parallel part of the polarization operator develops imaginary part $\Im \Pi_\|=\pi _ \|$~\cite{dgs2009, mv2009},
\be
\pi _ \| = -2 \a m ^ 2 I ^ 2 _ 1 (\gamma _ \t) \exp \l [ - \f {\pi m ^ 2} {eE}\r ],
\label{pipar}
\ee
with $\gamma _ \t = m \omega \sin \t / eE$ being the well-known Keldysh parameter and $\t$ is the angle between the field and the photon propagation direction. Finally, $I_n(z)$ is  the standard notation for the modified Bessel function. The  formula in Eq.(\ref{pipar}) is valid for arbitrary values of $\gamma _ \t$ as long as the photon energy is small as compared to the electron mass, $\omega^2 \ll m^2$.

The orthogonal part of the polarization tensor develops imaginary part only in the next order in $eE / m ^ 2$. We find here the following explicit expression for $\Im \Pi_\bot = \pi_\bot$ in this order in $eE / m ^ 2$, at arbitrary $\gamma_\t$ and small $\omega$:
\be
\pi _ \bot = -\f {\a} {\pi} e E \l ( 1 - I ^ 2 _ 0 (\gamma _ \t)\r )\exp \l [ - \f {\pi m ^ 2} {eE}\r ].
\label{pio}
\ee

We also argue here that there is another potentially important contribution to the $\pi _ \bot$ coming from the interaction of the photon with the electron magnetic moment
\be
\pi ^ m _ \bot = - \f {\a} {4}\, (\omega \, \sin \t)^2 \, I ^ 2 _ 1 (\gamma _ \t) \exp \l [ - \f {\pi m ^ 2} {eE}\r ]
\label{pim}
\ee
The suppression of this term in comparison with the parallel part (\ref{pipar}) is governed by the parameter $\omega^2/m^2$, rather than $eE/m^2$ as is the case for the term in Eq.(\ref{pio}).
Thus the magnetic contribution is more important if $\omega^2 \gg e E$, which situation is quite compatible with both $\omega^2$ and $e E$ being much smaller than $m^2$.

Our treatment is based on considering the semiclassical tunneling trajectory for the electron in the electric field. Within this approach the parallel part (\ref{pipar}) and the magnetic one (\ref{pim}) arise as purely classical correlators of respectively the classical current and the current associated with the magnetic moment on an unperturbed semiclassical trajectory. The term (\ref{pio}) arises from small fluctuations, `zitterbewegung', orthogonal to the electric field around the same unperturbed trajectory. The results in the equations (\ref{pipar}) - (\ref{pim}) are derived assuming that the deformation by the photon of the trajectory is small. At large photon energy this approximation is no longer valid and the deformation of the tunneling path should be taken into account. In what follows we find the semiclassical trajectory for an arbitrary energy of the photon and calculate the action on the resulting trajectory by purely geometrical means. As usual, the (Euclidean) action $S$ describes the exponential factor $e^{-S}$ in the rate. Our result reads as 
\be
\pi_{\|,\bot} \sim \exp \l [ \gamma _ \t- \l ( \f {2 m ^ 2} {eE} + \f {e E} {2 m^2} \, \gamma ^ 2 _ \t  \r )  \, \arctan \f {2 m^2} {e E \, \gamma_ \t}  \r ]~
\label{pia}
\ee
and reproduces that found in Ref.~\cite{dgs2009}. 

The material in this paper is organized as follows. In Section 2 we describe the treatment of the problem in terms of Euclidean space effective action, and in Section 3 present a calculation of the parallel and orthogonal absorptive parts $\pi_\|$ and $\pi_\bot$ of the vacuum polarization due to the `electric' interaction of the electron. In Section 4 we consider the contribution of the electron magnetic moment to the effective action and calculate its contribution to the `magnetic' part of $\pi_\bot$. In Section 5 we consider the tunneling trajectory deformed by the incident photon at arbitrary energy and
calculate the exponential factor in the rate of the process. We summarize and discuss our results in Section 6, and some technical points of the calculation are described in the Appendices A and B.

\section{Semiclassical approach.}
To find the production rate of the electron-positron pairs in an external electric field we can employ the method
similar to the one used in the problem of a metastable vacuum decay \cite{Coleman,vko,cc,Voloshin:1985id}. The production rate is related to the imaginary part of the vacuum energy $\Gamma = - 2 \mathrm{Im} E _ {\rm vac} $, which can be found in terms of the (Euclidean) path integral ${\cal Z}$ in the theory as
\be
e ^ {- E _ {\rm vac} T} = {\cal Z} \equiv \int \mathcal{D} \psi \, \mathcal{D} \bar { \psi } \mathcal{D} A _ \m e ^ { - S [\psi, A _ \m]}.
\ee
The contribution of configurations with $e^+e^-$ pairs, giving rise to the instability of the vacuum in the presence of an external electric field,  can be rewritten in terms of an integral over  closed trajectories of the electron \cite{Affleck,Dunne:2005sx,Dunne:2006st}.  The actual integration is then done semiclassically using the saddle point method. The stationary configuration (trajectory) for the action in this case is called the bounce and is given by a circle of a fixed radius $R =  m / (eE)$. The leading exponent and the determinant around the bounce combine to give the semiclassical result for the process rate per unit time and unit volume $V$: 
\be
\f {\Gamma} {V} = 2 \mathrm{Im} \int \mathcal{D}\gamma e ^ {-S[\gamma]} = \f {\l ( eE \r ) ^ 2} {4 \pi ^3} \exp \l [ - \f {\pi m ^ 2} {eE}\r ].
\label{gamma}
\ee
The imaginary part of the path integral arises from the integration over one negative mode of the action at the bounce configuration as explained in great detail in Refs.~\cite{Coleman,cc}. 

For the case at hand we use the same approach, except that we calculate the correlator of currents rather than the vacuum energy. The interaction between the electromagnetic field and the current is described by the familiar Lagrangian $\mathcal{L} _ {int} = A _ \m j _ \m$, so that the vacuum polarization operator is given by the standard expression 
\be
\Pi _ {\m \n} (x) =  \langle j _ \m (x) j _ \n (0) \rangle = {1 \over {\cal Z}} \,\int \mathcal{D}A \, \mathcal{D}\psi \, \mathcal{D} \, 
\bar {\psi} \, j _ \m (x) j _ \n (0) \, e ^ {-S[A,\psi]}.
\ee
Due to the instability (metastability) of the vacuum in an external electric field, the vacuum polarization develops an imaginary part $\pi(x) = \Im \Pi(x)$. In a semiclassical approach this imaginary part is evaluated similarly to the imaginary part of the vacuum energy by calculating the correlator of the currents on the bounce configuration. In terms of the effective action for the electron trajectories $\gamma$ the correlator then can be found by integrating over $\gamma$ the product of the currents on each of the trajectories:
\be
\langle j _ \m (x) j _ \n (0) \rangle = \int \mathcal{D}\gamma \, j ^ \gamma _ \m (x) j ^ \gamma _ \n (0) e ^ {-S[\gamma]}~,
\label{jjcorrd}
\ee
where the explicit form of the action $S[\gamma]$ in the external potential $A_\m^{\rm ext}$ has the standard form
\be
S = \oint _ \gamma \l ( m \sqrt {\dot {X} _ \m ^ 2} - e A _ \m^{\rm ext} \dot {X} _ \m \r ) ds,
\label{act}
\ee
with $X_\m(s)$ describing the trajectory $\gamma$ in terms of the position $X_\m$ as a function of the length parameter $s$, and the dot standing for the derivative $\dot X = dX/ds$.

It should be noted that the vacuum in a constant electric field is invariant under space-time translations. As a result the probability of the spontaneous pair creation is proportional to the the space time volume. This is reflected in Eq.(\ref{gamma}) in that a finite physical quantity is the rate per unit time and per unit volume. In terms of a path integration around the bounce configuration the translational invariance corresponds to the existence of four translational zero modes of the action at the bounce. The integration over these modes then corresponds to an integral over the space-time position $(x_\m)_B$ of the bounce, and the proper measure for the imaginary part of the integral (arising from the exponential $\exp(-S_{B})$ and the integration over the nonzero modes) can be read directly off Eq.(\ref{gamma}). For the bounce contribution to the imaginary part $\pi_{\m \n}$ of the vacuum polarization  $\Pi_{\m \n}$ the proper expression takes the form
\be
\pi_{\m \n} = -{\Gamma \over 2 V} \, \int \,\langle j _ \m (x) j _ \n (0) \rangle _{x_B} \, d^4 x_B= - {(e E)^3 \over 8 \pi^3} \, \exp \l [ - \f {\pi m ^ 2} {eE}\r ] \int \,\langle j _ \m (x) j _ \n (0) \rangle _{x_B} \, d^4 x_B~,
\label{intxb}
\ee
where the $\langle \ldots \rangle_{x_B}$ stands for the correlator calculated with a fixed position $x_B$ of the bounce.

\begin{figure}[ht]
  \begin{center}
    \leavevmode
    \epsfxsize=7cm
    \epsfbox{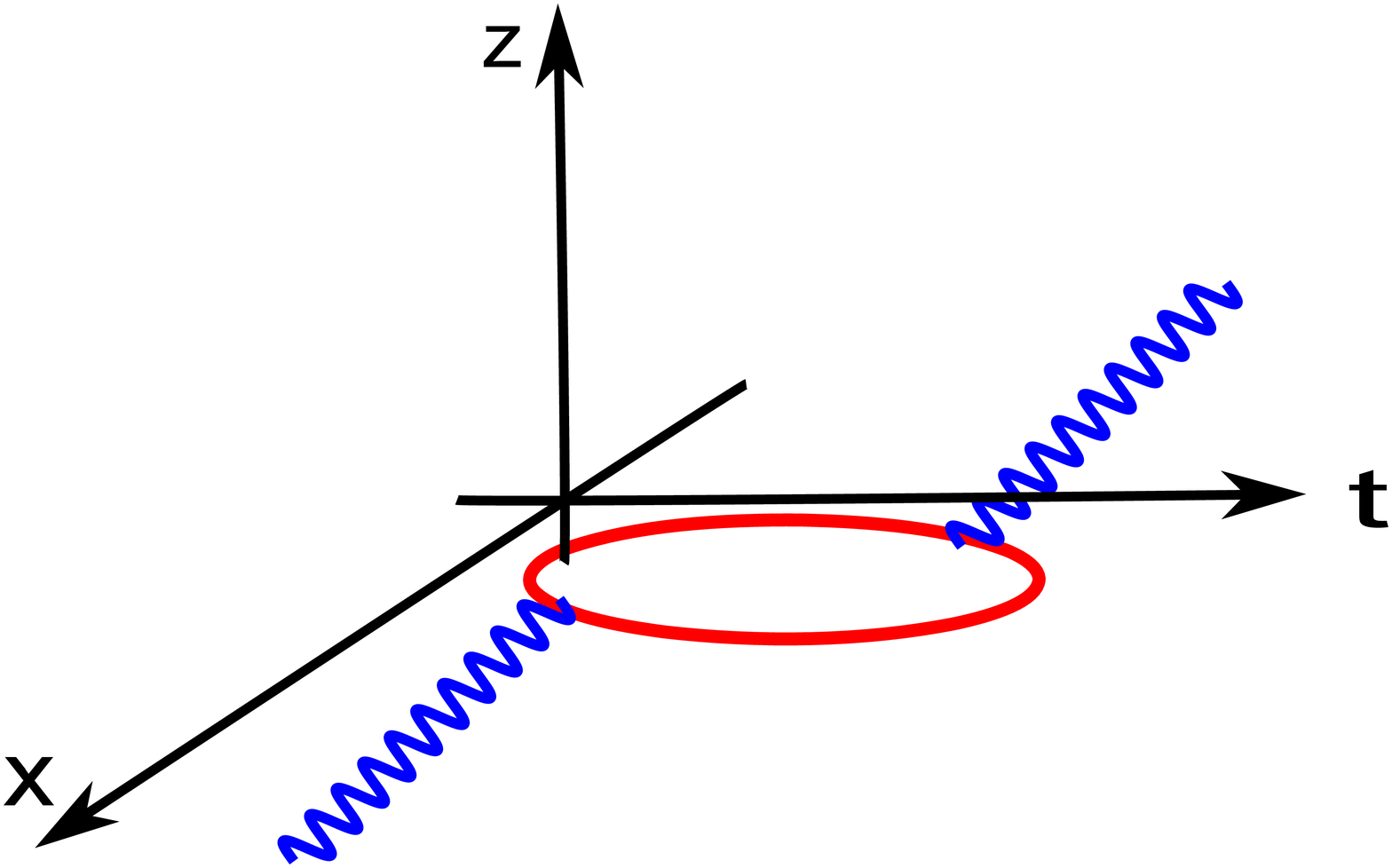}
    \caption{}
  \end{center}
\label{induced}
\end{figure}

\section{Calculating  $ \pi _ \|$ and $\pi _ \bot$.}
In this and the next section we consider the situation when the energy of the photon is small compared to the mass of the electron $\omega \ll m$. This limit allows one to use the same bounce configuration as in the spontaneous pair production case. To find the polarization operator in the leading order in the small parameter $eE / m ^ 2$ one can neglect all the spinor structure of the current $j _ \m ^ \gamma $ and take it in the form
\be
j _ \m ^ \gamma (x) = e \, \int \, \dot {X} _ \m (s) \delta^4 (x - X (s)) \, ds~.
\label{currg}
\ee
In the saddle point approximation one finds the tunneling trajectory, the bounce, which is a solution to the classical equations of motion corresponding to the action (\ref{act}) $X _ \m ^ B (s)$. In order to take into account small fluctuations around the stationary path one can parametrize the trajectory $X (s)$ as
\be
X _ \m (s) = X _ \m ^ B (s) + \xi _ \m (s).
\ee
Therefore the correlator (\ref{jjcorrd}) takes the form
\bea
&& \langle  j _ \m (x) j _ \n (0) \rangle = \nonumber \\
&& \! \!  \! \! \! \! e^2 \, \int \mathcal{D} \xi \, d s _ 1 \, d s _ 2 \,
\l [ {\dot X} ^ B _ \m (s _ 1) {\dot X} ^ B _ \n (s _ 2) +
{\dot \xi} ^ B _ \m (s _ 1) {\dot \xi}^ B _ \n (s _ 2) \r ] \delta^4 \l (x -  X ^ B (s _ 1) \r )
\delta^4 \l (X ^ B (s _ 2) \r ) e ^ {-S _ B - \delta S[\xi]}~.
\label{jjcorrsp}
\eea
The first term in the straight braces in (\ref{jjcorrsp}) corresponds to the calculation of the correlator using undisturbed electron trajectory (see Fig. 1), while the second term corresponds to the fluctuations of the bounce shown in Fig. 2.
\begin{figure}[ht]
  \begin{center}
    \leavevmode
    \epsfxsize=7cm
    \epsfbox{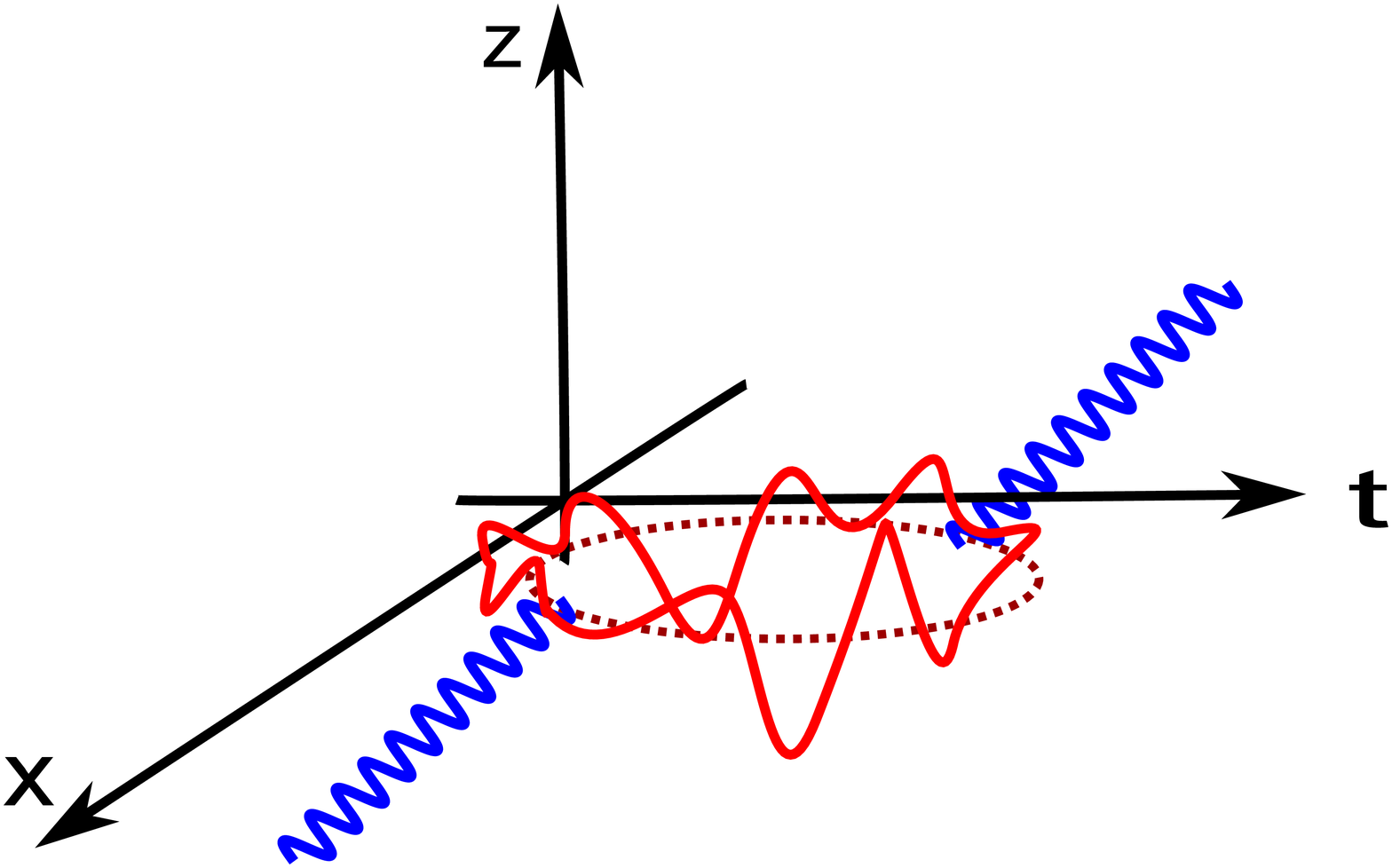}
    \caption{}
  \end{center}
\label{induced_pert}
\end{figure}

We choose the coordinate axes in such a way that the external electric field is in the $x$ direction, so that the bounce is a circle in the $(t,x)$ plane, and start with calculating the leading part in the correlator (\ref{jjcorrsp}). For a bounce centered at the origin, the current can be expressed in the cylindrical coordinates $(r,\varphi,y,z)$ as
\be
j _ \m ^ B (x) = e\, \int \, \dot {X} _ \m (s) \, \delta^4 (x - X (s)) \, ds = e n _ \m (\t) \delta (r - R) \delta (y) \delta (z),
\label{current}
\ee
where $n _ \m = ( - \sin \varphi, \cos \varphi, 0,0)$ is a tangential unit vector to the circle. As previously  argued, the leading contribution to the polarization tensor comes from the parallel part. Indeed, the current from the bounce has no components in the transverse to the plane $(t, x)$ directions. Therefore it is obvious that only
$j ^ B_ a j ^ B_ b$ with $a,b=0,1$ produce nonzero result. Integration over the zero modes $x _ B$ (the position of the bounce) can be readily performed giving
\be
\int d ^ 4 x_B \, j ^ B _ a (x) j ^ B _ b (0) = e ^ 2 \, \f {4 R^2 r _ a r _ b - \delta _ {a b} r ^ 4} 
{r ^ 3 \, \sqrt{4 R ^2 -r ^ 2}} \delta (y) \delta (z),
\label{jjint}
\ee
where $r$ is the length of the vector in the plane $(t, x )$, $r = \sqrt{t ^ 2 + x ^ 2 }$ (some details of the calculation can be found in Appendix A). Performing also the Fourier transform of the polarization operator
\be
\pi _ {a b} (q) = \int d ^ 4 x \, \pi _ {a b} (x) \, e ^ {i q x}~ 
\ee
and inserting the proper integration measure from Eq.(\ref{intxb}) one finally finds
\be
\pi _ {a b} (q ^ 2) = - 2 \a m^2 I _ 1 ^ 2 \l (\sqrt{-q^2} \, R \r ) \l ( \delta _ {a b} - \f {q _ a q _ b} { q  ^ 2 } \r)
\exp \l ( - \f {\pi m ^ 2} { e E } \r ),
\label{piab}
\ee
where $q ^ 2 = q _ 0 ^2 + q _ 1 ^ 2$. Clearly, the dependence on $\sqrt{-q^2} \, R$ in terms of the Euclidean variables translates into the dependence on $\gamma_ \t$ for an on-shell photon propagating in the Minkowski space. Indeed for a photon with four-momentum $k$ one has $-q^2 = \omega^2 - k_x^2 = \omega^2 \, \sin^2 \t$, so that
\be
\sqrt{-q^2} \, R \to \gamma_\t~,
\ee
given that $R = m/(e E)$. With this substitution 
the expression in (\ref{piab}) reduces to that in Eq.(\ref{pipar}), which coincides with the recently obtained result \cite{dgs2009, mv2009}.

The calculation of the second term in (\ref{jjcorrsp}) can also be done in a standard way. It is obvious that the contribution to the orthogonal part of the polarization tensor comes from the fluctuations of the trajectory in transverse directions, and for the purpose of calculation it is sufficient to consider one of the orthogonal directions, e.g. that along the $z$ axis. It then follows from (\ref{jjcorrsp}) that the orthogonal part of the polarization operator is determined by the path integral
\be
\Pi _ \bot = \langle j _ 3 (x) j _ 3 (0) \rangle = e^2 \,
\int \mathcal{D} \xi  \, ds_1 \, ds_2 \, {\dot \xi}^ B _ 3 (s _ 1)  {\dot \xi}^ B _ 3 (s _ 2) \delta^4 \l (x -  X ^ B (s _ 1) \r )
\delta^4 \l (X ^ B (s _ 2) \r ) e ^ {-S _ B - \delta S[\xi]} ~.
\label{xi3}
\ee 
Expanding the $\xi _ 3 (s)$ in a Fourier series, integrating over the amplitude of each mode
and making the Fourier transformation we find the expression (\ref{pio}), with the details of the calculation described in Appendix B.

\section{The magnetic moment term $\pi ^ m _ \bot$}
As it was mentioned before, there is another contribution to the orthogonal vacuum polarization $\pi _ \bot$ that is suppressed by the parameter $\omega^2 / m^2$ rather than by $(e E/m^2)$. This contribution arises from  the spinor structure of the current. The electron has a magnetic moment $\m =e/ (2 m)$. The interaction between the field and the magnetic moment in the rest frame of electron is given by $- \vec {\m} \vec {B}$. The relativistic expression for such an interaction has the form 
\be
\mathcal{L} _ {int} = \f {1} {4 m} \, \epsilon _ {\m \n \lambda \sigma} F _ {\m \n} f_ \lambda j ^ \gamma _ \sigma = 
\f {1} {2 m} \, \epsilon _ {\m \n \lambda \sigma} \p _ \m A _ \n  \m _ \lambda j ^ \gamma _ \sigma \to
- \f {1} {2 m} \, A _ \n \p _ \m \epsilon _ {\m \n \lambda \sigma} f _ \lambda  j ^ \gamma _ \sigma,
\label{mulor}
\ee
where $f_\lambda$ is a relativistic generalization of a unit vector in the direction of polarization of the electron. Namely, if $\vec f$ with $|\vec f|=1$ describes the polarization of the electron in its rest frame, in a frame where the electron four-momentum is $(\epsilon, \vec p)$ the four-vector $f_\lambda$ has the form: $[(\vec p \cdot \vec f)/m, \vec f+ \vec p \, (\vec p \cdot \vec f)/(\epsilon+m) m ]$.  The current $j ^ \gamma _ \sigma$ in Eq.(\ref{mulor}) is the same as given by the expression (\ref{currg}). Thus a calculation of $\pi ^ m _ \bot$ is reduced to an evaluation of the correlator of the currents
\be
j ^ m _ \n = - {1 \over 2m} \, \epsilon _ {\m \n \lambda \sigma} f_ \lambda \, \p _ \m  j ^ \gamma _ \sigma.
\label{mcurr}
\ee
The required expression can in fact be found using the formula for the correlator of the currents from Eq.(\ref{currg}). Indeed, the correlator can be reduced to the calculation of  $\langle j ^ \gamma _ a (x) j ^ \gamma _ b (0) \rangle$ as
\be
\langle j ^ m _ \n (x) j ^ m _ \rho (0) \rangle =  { 1 \over 4 m^2} \,
\varepsilon _ {\m \n \lambda \sigma } \,  f_ \lambda \, \varepsilon _ {\a \rho \beta \tau } \,  f _ \beta \, 
\p _ \m \, \p _ \a \, \langle j ^ \gamma _ \sigma  (x) j ^ \gamma _ \tau (0) \rangle.
\ee
We then find that after averaging over the transverse indices 2 and 3 the Fourier transform of this expression takes a simple form:
\be
\pi^m_\bot={1 \over 2} \, (\pi_{22} + \pi_{33}) \, = -{1 \over 8 m^2} \, f_\m^2 \, q^2 \pi_\| = -{1 \over 8 m^2}  \, q^2 \, \pi_\|(q^2)
\label{pimpar}
\ee
with $q^2$, as previously, being the square of the two dimensional Euclidean vector related to the photon momentum as $-q^2 = \omega^2 \, \sin^2 \t$. Clearly, the formula (\ref{pimpar}) 
gives the relation between the expressions in Eq.(\ref{pim}) and Eq.(\ref{pipar}).

\section{Polarization operator at arbitrary $\omega$.}
In order to find the vacuum polarization tensor for arbitrary frequency of the photon one should take into account the deformation of the tunneling trajectory by the photon energy and momentum. This deformation is driven by two opposing factors. On one hand the photon supplies the energy to the pair while on the other hand it transfers its momentum to the $e^+e^-$ pair which increases the energy barrier for tunneling. A proper treatment of these effects  amounts to finding the new configuration and calculating the action on it.
\begin{figure}[ht]
  \begin{center}
    \leavevmode
    \epsfxsize=7cm
    \epsfbox{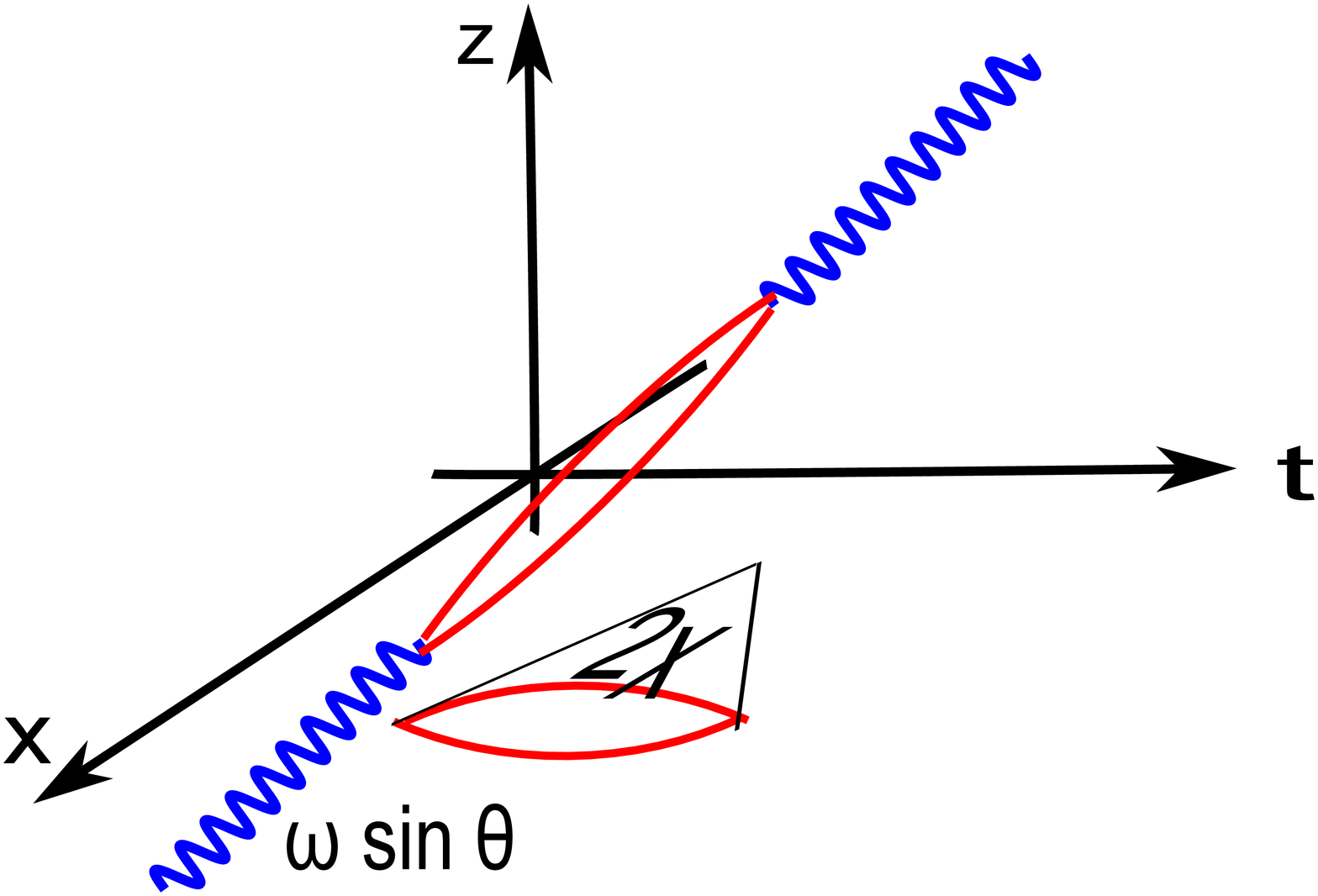}
    \caption{}
  \end{center}
\label{induced_deformed}
\end{figure}
The notation takes a simple form if one first performs a Lorentz transformation in the direction of the electric field such that sets the longitudinal momentum of the photon to zero,  $k_x=0$, and also the transverse axes are chosen so that $k_y=0$. In terms of the initial energy $\omega$ and the incident angle $\t$ of the photon the four-momentum of the photon in the new frame has the form $(\omega \sin \t,\, 0,\, 0, \omega \sin \t)$. The proper trajectory is then found by solving the classical (Euclidean) equations of motion corresponding to the modified action
\be
S = m \int \sqrt { {x'} _ \m ^ 2 } d \varphi - e \int A _ \m  { x' } _ \m d \varphi - T \, \omega \sin \t  + Z \, \omega \sin \t  , 
\label{newact}
\ee
where $\varphi$ is the angular coordinate of the projection of the trajectory on the $(t,x)$ plane, and $x'=d x/d \varphi$. The expression (\ref{newact})
allows for the electron to have nonzero momentum in the $z$ direction (it should be mentioned that in the Euclidean version this momentum is imaginary $i p $). $T$ and $Z$ are the sizes of the loop along the $t$ and $z$ directions correspondingly, and the last two terms in (\ref{newact}) are given by the negative of the Euclidean action of the photon, corresponding to the gap in the photon propagation over the duration and the extent in the $z$ direction of the 
$e ^ + e ^ -$loop.  Such an approach is similar to the one  described in \cite{SV1985,m2005}. As usually, the equations of motion for the particle in external electromagnetic field are
\be
m \, \f {d} {d\varphi} \f {  { x' } _ \m } {\sqrt {{x'} ^ 2 }} = - e F _ {\m \n}  { x' } _ n.
\ee
Solving these one gets
\bea
t & = & \f {\sqrt {p ^ 2 + m ^ 2}} {eE} \cos {\varphi} + c _ 0, \nonumber \\
x & = & \f {\sqrt {p ^ 2 + m ^ 2}} {eE} \sin {\varphi}+ c _ 1, \nonumber \\
z & = & \f {p} {e E} \varphi + c _ 3,
\label{ttraj}
\eea
where $c_\m$ are constants. One can see from the expressions (\ref{ttraj}) that the projection of the trajectory on the $(t,x)$ plane consists of two arcs of a circle with the radius $ {\sqrt {p ^ 2 + m ^ 2}} / {eE}$. Introducing the angle spanned by each of the arcs as $ 2 \chi$ and substituting the solution back to Eq.(\ref{newact}), one finds
\be
S = \f {4 m^ 2 \chi} {eE} - \f {2 \chi (m^ 2 + p^2) } {eE} + \f {m^ 2 + p^2} {eE} \sin 2 \chi - 
\f {2 \omega \, \sin \t \, \sqrt {p ^ 2 + m ^ 2}}{eE} \sin \chi + \f {2 \omega \, \sin \t \, p \chi} {eE}.
\ee
Minimizing the action with respect to $\chi$ and $p$ gives the expression (\ref{pia}), which coincides with the one derived by another technique in \cite{dgs2009}~\footnote{One can readily notice that the derived in this way values of $\chi$ and $p$ correspond to enforcing the energy and momentum conservation in the creation and annihilation of the pair by the photon}.

\section{Summary and discussion.}
In this paper we have considered the photon-stimulated Schwinger process in terms of a calculation of the vacuum polarization operator by purely semiclassical means. At the value of the electric field much smaller than $E_{\rm crit}$ the radius $R=m /(eE)$ is large in comparison with the Compton wavelength $1/m$, so that an expansion in $e E/m^2$ is in fact an expansion in the curvature of the semiclassical trajectory divided by $m^2$. A small curvature at each point of the tunneling path implies that one can employ a nonrelativistic expansion in the co-moving frame, which in fact is a way of interpreting the treatment described in this paper.

The dominant in the parameter $eE/m^2$ effect arises for the photons with polarization along the external electric field. As can be seen from the discussed here calculation this effect is described by essentially the classical current of the charged particles on the tunneling trajectory interacting with field of the incident photon. Such an origin of this leading contribution is also in full agreement with the behavior described in our recent paper~\cite{mv2009} where the same effect originates from a classical interaction in a thermal treatment of the Schwinger process. Clearly, this part is not sensitive to the electron spin and has exactly the same form for creation of pairs of spinless particles (modulo the statistical spin factor of two). In terms of a nonrelativistic expansion in the co-moving frame this contribution is that of the electrostatic interaction.

For photons with polarization orthogonal to the electric field, however, there are two types of contribution to the stimulated pair production. One source, also independent of the spin, arises from the interaction of the incident photon with the current due to the fluctuations of the tunneling trajectory in the direction perpendicular to the external field. This effect, described by Eq.(\ref{pio}), in fact requires an equivalent of a one-loop calculation as discussed in the Appendix B. In the nonrelativistic interpretation this contribution is that of the interaction of a particle having velocity $\vec v$ with the vector potential $\vec A$. It can be noted in connection with the calculation of this effect that the photon energy dependence required by the gauge invariance is ensured by the subtraction in Eq.(\ref{pio}) of the constant term, which automatically arises as a result of the summation in Eq.(\ref{pi33f}). This constant term is analogous to the well known `sea gull' type graph in QED of scalar particles.  

An effect of the electron spin in the discussed process arises for photons with polarization orthogonal to the external field through the interaction in the co-moving frame of the electron magnetic moment with the photon. As discussed in the Section 4, once the `quantum' origin of this contribution is absorbed in the magnetic interaction, the rest of the calculation is again essentially classical.
Formally, the effect of this term is of a higher (second) order in the expansion in the ratio $eE/m^2$ as compared to that given by Eq.(\ref{pio}), and is suppressed at small values of $\gamma_\t$. However at $\gamma_\t$ of order one and larger the ratio of the magnetic contribution to that in Eq.(\ref{pio}) is of the order of $(\omega \, \sin \t)^2/(e E)$, which is an independent dimensionless parameter in the problem, and there is generally no reason to assume that this parameter is small.

A deformation of the tunneling trajectory due to the interaction with the incident photon can be neglected as long as the photon energy is small: $\omega \ll m$. In terms of the expansion in the co-moving frame this corresponds to appearance of only a weak `kink' in the trajectory at the point of interaction with the photon, which does not invalidate the expansion in small curvature. However at larger $\omega$ the kink is `strong' and the approximation of an unperturbed tunneling path is not valid. Although at present a full calculation in this situation is not available, the exponential factor in the stimulated rate of pair creation can be found using the effective action approach, previously applied~\cite{SV1985,m2005} in similar problems related to an induced decay of metastable vacuum. Proceeding in this way, we have arrived at the exponential factor in Eq.(\ref{pia}), which is valid at an arbitrary energy $\omega$ of the photon. The exponential power in this factor can be expanded in $\gamma_\t$ as
\be
\pi_{\|,\bot} \sim \exp \l \{ - {\pi \, m^2 \over e E} + 2 \, \gamma_\t - {\pi \over 4}\, \gamma_\t^2  \, { e E \over  m^2} + O \l [ {\gamma_\t^3 \, (e E)^2 \over m^4 } \r ] \r \}~,
\label{piam}
\ee
which shows that indeed the parameter for the expansion is $\gamma_\t \, e E/ m^2 = \omega \sin _\t/m$, and the higher terms are small as long as $\omega \ll m$. It is also satisfying to notice that the linear in $\gamma_\t$ term in Eq.(\ref{piam}) agrees with the large $\gamma_\t$ asymptotic behavior of the Bessel functions in the equations (\ref{pipar}) - (\ref{pim}). In other words, those expressions describe, including the preexponential factors, the matching between the low energy  and the high energy behavior of the pair creation rate. Clearly the overlap region for these two types of expressions is determined by the condition $1 \ll \gamma_\t \ll m^2/(eE)$.

\section*{Acknowledgments}

The work of
A.M. is supported in part by the Stanwood Johnston grant from the Graduate
School of the University of Minnesota, RFBR Grant No. 07-02-00878 and by the
Scientific School Grant No. NSh-3036.2008.2.
The work of M.B.V. is supported in part by the DOE grant DE-FG02-94ER40823.

%\newpage
\section*{Appendices}

\appendix

\section{$\pi_\|$}

%\subsection{Appendix A. \label{A}}
Let us consider the circular electron trajectory (the bounce) with the center  located at $x_B=(t_ B,x_B,y_B,z_B)$. The current at points $(0,x,y,z)$ and $(0,0,0,0)$ (we can always choose such a coordinate system) is found using the expression (\ref{current})
\bea
j ^ B _ a (x) & = & e \, \f {(x _ B - x, -t _ B)} {r _ 2} \, \delta (r _ 2 - R) \, \delta (y - y _B) \,
\delta (z - z_ B), \nonumber \\   
j ^ B _ b (0) & = & e \, \f {(x _ B , -t _ B)} {r _ 1} \, \delta (r _ 1 - R) \, \delta (y _ B) \, \delta (z _ B),
\eea
where $r_ 1 = \sqrt {t _ B ^2 +x _ B ^ 2}$, $ r _ 2 = \sqrt {t _ B ^2 +(x _ B - x) ^ 2}$, and $a$ and $b$ are the two-dimensional indices in the $(t,x)$ plane. Therefore one finds that the l.h.s. of the (\ref{jjint}) can be rewritten as
\bea
e ^ 2 \, \int \f {d t _ B \, d x _ B} {r _ 1 r _ 2}
\l (
\begin{array}{ccc}
(x _ B - x) x _ B & - (x_ B - x) t _ B \\
-x _ B t _ B & t _ B ^2
\end{array} 
\r )
\delta (r _ 1 - R)\delta (r _ 2 - R) \delta (y)\delta (z).
\eea
One can integrate over $r _ 1$ and $r _ 2$ instead of $t _ B$ and $x _ B$, taking into account the Jacobian of the transformation, 
\be
\l | \f {\p (t _ B, x _ B)} {\p (r _ 1, r _ 2)}\r | = \f {r _ 1 r _ 2} {x |t _ B|}.
\ee

The integration of the delta functions reduces to calculation of the expression in the integral for two configurations shown in Fig.~4.
\begin{figure}[ht]
  \begin{center}
    \leavevmode
    \epsfxsize=7cm
    \epsfbox{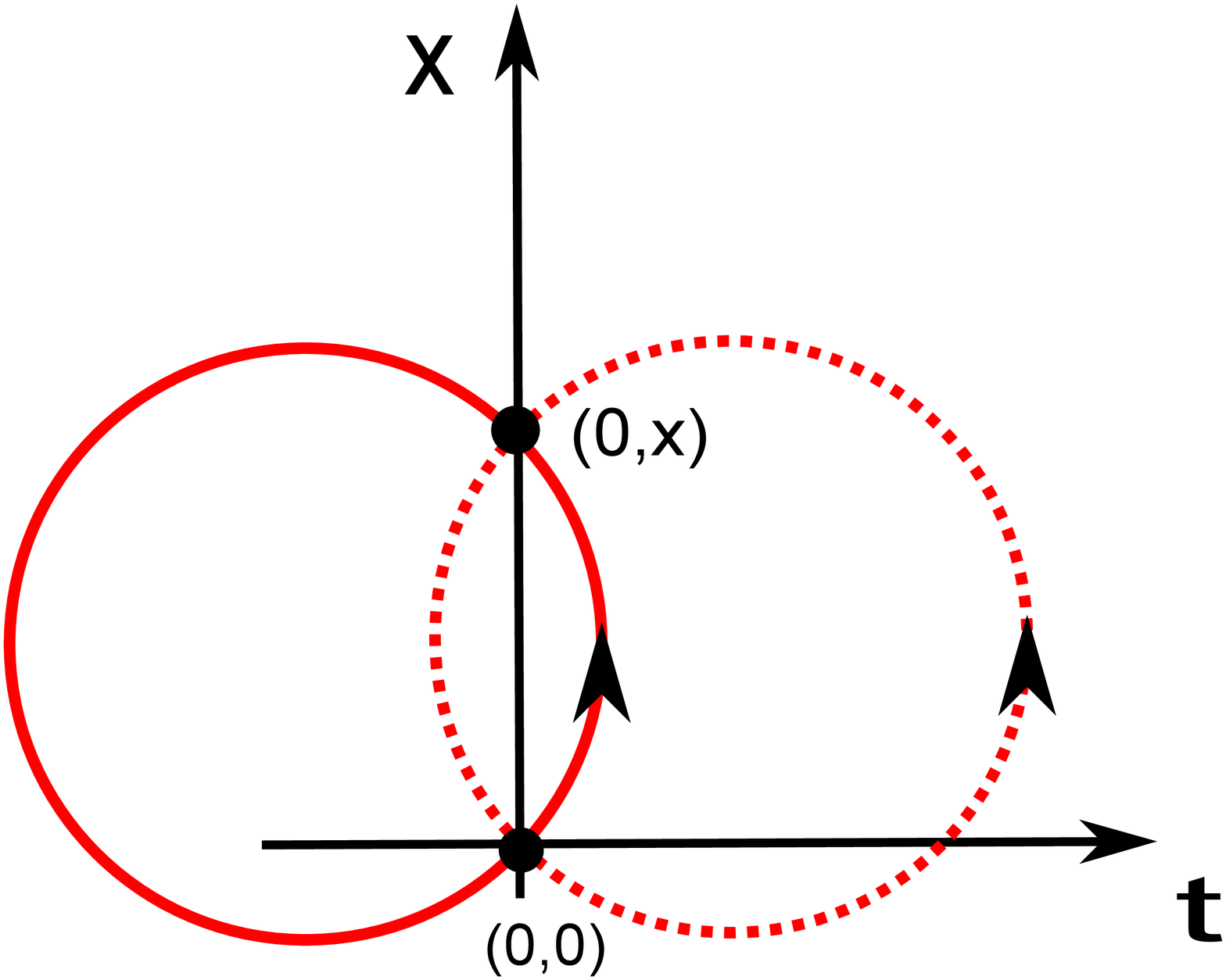}
    \caption{}
  \end{center}
\label{2_circles}
\end{figure}
As a result one gets the l.h.s. of Eq.(\ref{jjint}) in the form
\be
e ^2
\l (
\begin{array}{cc}
-x ^2 & 0 \\
0 & 4 R ^2 -x ^ 2
\end{array}
\r)
\f {\delta (y)\delta (z)} {x \sqrt {4 R ^2 - x ^2}}.
\label{jjintps}
\ee
As follows from the symmetries in the problem the integral can be generally written in the form
\be
\int d ^ 4 x_B \, j ^ B _ a (x) j ^ B _ b (0) = r _ a r _ b B + A \delta _ {a b},
\label{jjanz}
\ee
where yet unknown coefficient functions $A$ and $B$ depend only on the SO(2) invariant length $r = \sqrt {t ^ 2+ x  ^2}$.
Comparing (\ref{jjintps}) with (\ref{jjanz}) we find these coefficient functions and arrive at the following expression
\be
\int d ^ 4 x_B \, j ^ B _ a (x) j ^ B _ b (0) = e ^ 2 \, \f {4 R^2 r _ a r _ b - \delta _ {a b} r ^ 4} 
{r ^ 3 \, \sqrt{4 R ^2 - r ^ 2}} \delta (y) \delta (z).
\ee
As a cross check it can be readily verified that the found expression satisfies the current conservation $\p_\mu \langle j  _ \mu (x) j _\nu (0) \rangle =0$.
Due to this property the  Fourier transform of this correlator has the standard orthogonal form
\be
\pi _ {a b} (q ^ 2) =  \l ( \delta _ {a b}- {q _ a q _ b \over q^2}\r ) \, \pi (q ^ 2),
\ee
where $\pi (q ^ 2) = {\pi _ { a a } (q ^ 2) } $ (the summation over $a$ is implied).
\bea
\pi (q) & = & \f {1}{2} \, \f {\Gamma} {V} \, e^2 \, \int \f {4 R^2 -2 r ^2 } {\sqrt {4 R ^ 2 - r ^ 2}}\delta (y) \delta (z)
e ^ {i q r \cos \varphi} r \, d r \, d \varphi \, dy \, dz \nonumber \\ 
& = & 2 \pi e ^2 \, \f {1}{2} \, \f {\Gamma} {V} \,
\int \f {4 R^2 -2 r ^2} {\sqrt {4 R ^ 2 - r ^ 2}} J _ 0 (q r) \, d r \nonumber \\
& = & \f {1}{2} \, \f {\Gamma} {V} \, 4 \pi ^2 R ^ 2 e ^2 \, J _ 1 ^ 2(q R).
\eea
Taking into account that $q ^ 2 = - \omega ^ 2 \sin ^ 2 \t$, we find
the expression in Eq.(\ref{pipar}) for $\pi_\|$.

\section{The `electric' part of $\pi_\bot$}

%\subsection{Appendix B.}
The part of the action $\delta S[\xi]$ arising from the fluctuations of the electron trajectory in the transverse direction $\delta z= \xi_3$ has the form
\be
\delta S[\xi]  = S_3[\xi_3]- e \int A _ 3 \dot {\xi} _ 3 ds,
\ee 
with
\be
S_3[\xi_3]=\f {m} {2} \int ds \, \dot {\xi} _ 3 ^ 2~,
\ee
and
where $s$ is an arbitrary parameter on the trajectory, we choose it to be the length parameter of the unperturbed bounce, so that $s \in [0, 2 \pi R]$. For the bounce centered at the origin on can write the additional current associated with the fluctuation as
\be
 {j} ^ \xi _ 3 = e \, \dot{\xi_3} (s) \delta (r - R) \delta (y) \delta (z).
\ee
In order to find the correlator
\be
\langle { j } ^ \xi _ 3 (x)  { j } ^ \xi _ 3 (0) \rangle =
\int \mathcal {D} \xi  { j } ^ \xi _ 3 (x)  { j } ^ \xi _ 3 (0)  e ^ { - S_ B - S [\xi] },
\ee
we again choose the coordinate system where $x = (0, x, y,z)$. Then 
\bea
{ j } ^ \xi _ 3 (x) & = & e \,\dot{\xi} _ 3 (s_ 1) \, \delta (r _ 1 - R) \, \delta (y _ B - y) \, \delta (z _ B - z), \nonumber \\
{ j } ^ \xi _ 3 (0) & = & e \, \dot{\xi} _ 3 (s_ 2) \, \delta (r _ 2 - R) \, \delta (y _ B) \, \delta (z _ B).
\eea
According to Eq.(\ref{intxb}) the contribution of this current to the imaginary part of the polarization operator can then be written as
\be
\pi_{33}(x) = - \f {1}{2} \, \f {\Gamma} {V} \,
{\int \, d ^ 4 x _ B \, \mathcal{D} \xi _ 3 \, { j } ^ {\xi B} _ 3 (x) \,  { j } ^ {\xi B} _ 3 (0)   \, e ^ {- S _ 3} } /
{\int \mathcal{D} \xi _ 3 \, e ^ {- S _ 3} },
\ee
where the $j ^ {\xi B}$ is the current $j ^ {\xi}$ corresponding to the bounce centered at $x_B$. Integrating over the position of the bounce one gets
\be
\pi_{33}(x)  = - \f {1}{2} \, \f {\Gamma} {V} \, e^2 \,\f {4 R ^2} {x \sqrt {4 R ^ 2 - x ^ 2}} \, \delta (y) \delta (z) \, \int \mathcal{D} \xi _ 3 \, \xi _ 3 ( \bar {s}) \xi _ 3 ( - \bar {s}) 
e ^ {- S _ 3} / \int \mathcal{D} \xi _ 3 \, e ^ {- S _ 3},
\label{pi33xi}
\ee
where $\bar {s} = R \, \arctan (x/ \sqrt {4 R ^ 2 - x ^ 2})$. An expansion of the fluctuation $\xi_3(s)$ in the Fourier series
\be
\xi _ 3 (s) = \f {a _ 0} {\sqrt{2 \pi}} + 
\sum _ {n = 1} ^ \infty \l [\f {a _ n} {\sqrt{\pi}}\cos \f {n \, s } {R}+\f {b _ n} {\sqrt{\pi}}\sin \f {n \, s } {R} \r ],
\ee
reduces the expression (\ref{pi33xi}) to the following
\bea
\pi_{33}(x) & = & -\f {1}{2} \, \f {\Gamma} {V} \, e^2 \, \f {4 R ^2} { \pi x \sqrt {4 R ^ 2 - x ^ 2}}  \nonumber \\ 
& \times & \int [da \, db] \, \sum _ {n = 1} ^ \infty n ^ 2 \l ( b _ n ^ 2 \cos ^ 2 \f {n\bar{s}} {R} - a _ n ^ 2 \sin ^ 2 \f {n\bar{s}} {R} \r )
e ^ {- \sum _ { l } \f {m l ^ 2} {2 R} (a _ l ^ 2 + b _ l ^ 2 )} \nonumber \\
& \times & \l ( \int [da \, db] e ^ {- \sum _ { l } \f {m l ^ 2} {2 R} (a _ l ^ 2 + b _ l ^ 2 )} \r ) ^ {-1} \, \delta (y) \delta (z) \nonumber \\
& = &
- \f {1}{2} \, \f {\Gamma} {V} \, e^2 \,\f {4 R } {m \pi x \sqrt {4 R ^ 2 - x ^ 2}} \sum _ {n = 1} ^ \infty \cos \f {2 n \bar{s}} {R} \, \delta (y) \delta (z) \nonumber \\
& = & - \f {1}{2} \, \f {\Gamma} {V} \, e^2 \, \f {4 R } {2 \pi m  x \sqrt {4 R ^ 2 - x ^ 2}} 
\l [ \f {\pi} {2} \delta \l (\f {\bar {s}} {R} \r ) - 1 \r ] \, \delta (y) \delta (z).
\label{pi33f}
\eea
The component $\pi_{33}$ of the vacuum polarization is a scalar with respect to rotation in the $(t,x)$ plain. Therefore it is a function of the invariant $r=\sqrt{t^2+x^2}$, so that the invariant expression for the correlator in an arbitrary coordinate system can be obtained substituting $x \to r$ in Eq.(\ref{pi33f}). Proceeding in this manner and performing the Fourier transform, we find
\be
\pi _ {33}= \pi _ \bot = \f {\a} {\pi} e E \l (1 - I ^ 2 _ 0 (\gamma _ \t) \r )\exp \l [ - \f {\pi m ^ 2} {eE}\r ].
\ee

%\newpage

\end{document}